\documentclass[aps,amsmath,amsfonts,11pt]{revtex4}
\usepackage{graphicx}
\usepackage{epstopdf}
\usepackage{epsfig,graphicx}
\usepackage[english]{babel}
\usepackage{amsfonts}
\usepackage{amsmath}
\usepackage{latexsym}
\usepackage{graphics,bm}
\usepackage{dcolumn}
\usepackage{natbib}
\usepackage{bm}
\usepackage{rotating}

\begin{document}

\title{Optical response properties of hybrid electro-opto-mechanical system interacting with a qubit}

\author{ Tarun Kumar$^{1}$, Surabhi Yadav$^{2}$ and Aranya B Bhattacherjee$^{2}$ }

\address{$^{1}$Department of Physics, Ramjas College, University of Delhi, Delhi-110 007, India} \address{$^{2}$Department of Physics, Birla Institute of Technology and Science, Pilani,
Hyderabad Campus,  Telangana State - 500078, India}

\begin{abstract}
We investigate the optical response of a hybrid electro-optomechanical system interacting with a qubit. In our experimentally feasible system, tunable all-optical-switching, double-optomechanically induced transparency (OMIT) and optomechanically induced absorption (OMIA) can be realized. The proposed system is also shown to generate anomalous dispersion. Based on our theoretical results, we provide a tunable switch between OMIT and OMIA of the probe field by manipulating the relevant system parameters. Also, the normal-mode-splitting (NMS) effect induced by the interactions between the subsystems are discussed in detail and the effects of varying the interactions on the NMS are clarified. These rich optical properties of the probe field may provide a promising platform for controllable all-optical-switch and various other quantum photonic devices.
\end{abstract}

\maketitle

\section{Introduction}
\indent 
The technological advancements in micro and nano-mechanical oscillators have shown great potential in exploring novel quantum devices and demonstrating the transition from quantum to classical physics \cite{poot, blencowe}. By coupling mechanical resontors to other quantum objects, such a transition can be demonstrated. The quantum systems that have been coupled to mechanical resonators include optical cavities \cite{kiesel, kippenberg, schwab1}, electron spin\cite{chui}, superconductiong qubit circuits \cite{geller1, sun, schwab2, schwab3, geller2, semba, nori1, tian, roukes}, transmission line resonators \cite{bruder, schwab4, hertzberg, nori2, sill}, nitrogen vacancy centres, and quantum dots \cite{seidelin, lukin1, lukin2}. For example, phonon blockade mechanism demonstrates the quantization of mechanical oscillator\cite{nori3}. One of the challenging technologies in photonics is the manipulation of light for quantum information processing and optomechanical systems have played an important role in this direction\citep{painter, cleland, kippenberg2, simmonds, wang1, clerk}.  These systems exhibit optomechanically induced transparency (OMIT)\cite{laurat, weis, huang} which is an analogue of the well known electromagnetically induced transparency (EIT) in quantum optics. This phenomena is utilized to slow down or stop light\cite{painter2, wang2, wang3} and thus finds application in quantum communication and quantum information platforms. Hybrid quantum systems provide us opportunities for exploring many quantum physical phenomena, providing opportunities for quantum-enhanced devices\citep{painter, cleland, kippenberg2, simmonds, rabl, wang4, sill2, nori5, moore}.
In a semiconductor microcavity, the strong coupling between the cavity mode and exciton mode of a quantum dot(QD) or a quantum well (QW) leads to the formation of polaritons\cite{deppe, forchel, bloch}, which has a wide application in quantum information processing, quantum networking\citep{small, zeilinger} and optical switching\citep{vuckovic1, vuckovic2, waks}. Strongly coupled systems also exhibit the phenomena of normal mode splitting (NMS). NMS is the process of energy exchange between two nearly degenerate modes of the system and occurs when energy exchange between different modes of the system takes place at a rate faster than its dissipation to the surrounding\cite{vitali}. It has been shown that cooling of mechanical resonator in the resolved sideband regime at high pump power leads to the appearance of NMS \citep{dob}. NMS has also been shown to occur in system comprising of ultracold atoms in cavity optical lattice \citep{abb} and cavity optomechanics in a nonlinear optical cavity \citep{tarun,sonam}.  \indent\\
	A quantum interface transfers quantum state between different degrees of freedom and can be implemented using micro-mechanical oscillators. An example of such system is an electro-opto-mechanical system in which mechanical oscillator is coupled to a microwave as well as an optical cavity\citep{barz}. The existence of EIT in such a hybrid opto-electro-mechanical system has been predicted\citep{qu} A similar setup was used to show  the possible realization of controllable strong kerr nonlineraity even in the weak coupling regime\citep{lu}. Strong interactions between the electrical, mechanical and optical modes was demonstrated in a pizoelectric optomechanical crystal\citep{bochmann}. In an interesting experimental development, circuit cavity quantum electrodynamics was integrated with phonons\citep{pirk}. These hybrid electro-opto-mechanical systems enable the reversible conversion of quantum states between microwave and optical photons\citep{tian2}. A novel scheme was also proposed recently, which could generate a microwave-controllable optical double optomechanically induced transparency (OMIT) in a hybrid pizo-optomechanical cavity\cite{wu}. \indent\\
	In this work, we investigate a hybrid electro-opto-mechanical system in the presence of a two level system (Qubit) coupled to the mode of a mechanical resonator via the Jaynes-Cummings interaction. We explore the optical response properties such as optical bistability and output transmission as a function of the various interaction parameters. We demonstrate here that in the resolved side band regime, the system exhibits tunable optical switching behavior. In the nearly resolved side band regime (sideband resolved regime) \citep{wu}, the output transmission demonstrates the existence of double-optomechanically induced transparency (OMIT) along with anomalous dispersion (negative group velocity). In the resolved sideband regime, we observe optomechanically induced absorption (OMIA).  In particular, we are able to switch between OMIT to OMIA by manipulating the interactions and the optical cavity decay rate. The double-OMIT effect has been studied theoretically in numerous optomechanical systems \citep{qu,wu,wang5}. It was first theoretically shown that when the group velocity exceeds the light velocity in vacuum, the dispersion is anomolous \citep{brill}. In the past few years, cavity optomechanics has been shown to generate OMIT and slow light propagation \citep{kippenberg2,weis,huang,safavi,cyu}. Further, anomolous dispersion has been explored theoretically leading to some interesting results \citep{hwu,jqin,akram}. Experimental demonstration of fast and slow light have paved the way towards applications in telecommunications, interferometry and quantum-optomechanical memory \citep{dechang}. Similar to the optical response effect of OMIT, OMIA have been investigated in numerous optomechanical systems \citep{qu,wang5,prakash,yzhang,twang,liao}.  In addition, we study the NMS to elucidate the physics of energy exchange between the various modes of the systems. Earlier studies on optomechanical device coupled to a two level system includes two color EIT\cite{wang5}, entanglement dynamics of optical and mechanical modes using a QD\citep{yuan}, coherent perfect transmission mediated by a qubit embeded in a hybrid optomechanical system\citep{yzhang, bhattacherjee}.

\section{Theoretical Framework}
	\indent 
	 The proposed hybrid electro-optomechanical system is shown in Fig.1. It is composed of a mechanical resonator (MR) which on one side is capacitively coupled to the microwave field of a superconducting microwave cavity (MC) and, on the other side is coupled to the field of an optical cavity (OC). In addition, the MR with frequency $\omega_{b}$ is coupled to a two-level defect (qubit) described by the Jaynes-Cummings Hamiltonian \citep{wang5,ramos}. The microwave resonator with a resonance frequency $\omega_{c}$ is driven by a strong field with amplitude $E_{m}$ and frequency $\omega_{mi}$, while the optical cavity with a resonance frequency $\omega_{a}$ is driven by a pump laser with amplitude $E_{p}$ and frequency $\omega_{ai}$. In addition, the optical response can be probed by a weak probe laser with amplitude $E_{pr}$ and frequency $\omega_{pr}$. 
	 
The total Hamiltonian of the system in the rotating frame of pump laser can be written as-

\begin{eqnarray}\label{equation1}
H&=& \hbar\delta_aa^{\dagger}a+\hbar\delta_cc^{\dagger}c+\hbar\omega_{b}b^{\dagger}b+(\hbar/2)\omega_{q}\sigma_{z}-\hbar{g_{om}}a^{\dagger}a(b+b^{\dagger})+\hbar{g_{em}}c^{\dagger}c(b+b^{\dagger})
\nonumber\\
&+&\hbar{g}(b^{\dagger}\sigma_{-}+b\sigma_{+})+\iota\hbar{E_p}(a^{\dagger}-a)+\iota\hbar{E_m}(c^{\dagger}-c)+\iota\hbar{E_{pr}}(a^{\dagger}e^{-\iota\delta{t}}-a{e^{\iota\delta{t}}})
\end{eqnarray}	
	
\begin{figure}
\includegraphics[scale=0.7]{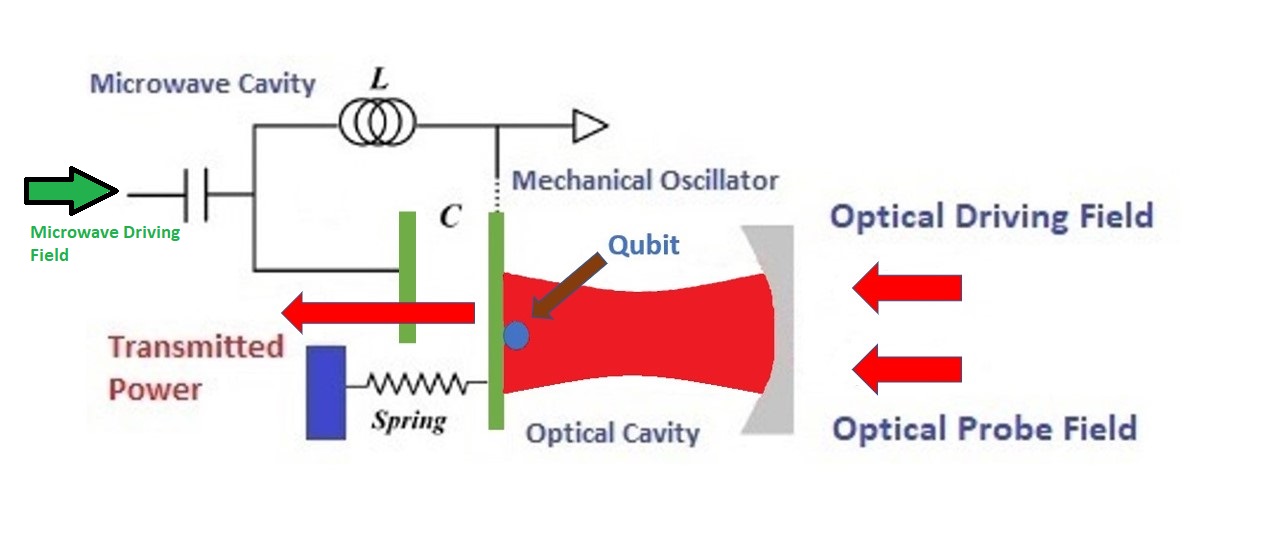}
\caption{Schematic diagram of the proposed  hybrid electro-optomechanical system. An embedded qubit is interacting with the mechanical oscillator. The mechanical oscillator also couples to both the microwave and optical cavity.}
\end{figure}

where $\delta_a = \omega_a-\omega_{ai}$ and $\delta_c = \omega_c-\omega_{mi}$ are the detunings of the optical field from the optical cavity and microwave field from the microwave cavity respectively. $\delta = \omega_{pr}-\omega_{ai}$ represents the detuning of the optical probe field from the pump field. In equation (1) of the Hamiltonian, first and the second terms represent the free energies of the optical mode and the microwave mode respectively. The third term gives the free energy of the movable mirror. The fourth term represents the energy of the the semiconductor quantum dot. Here $\omega_q$ is the transition frequency between two levels of the QD. Also $\sigma_-$ and $\sigma_+$ are the lowering and raising operators respectively of the two level QD. The fifth term represents the optomechanical coupling between the cavity mode and the mechanical mode where $g_{om}$ being the single photon optomechanical coupling constant, given by $(\omega_a/L\sqrt{\hbar/m\omega_{b}})$, m being the effective mass of the mechanical mode and L being the effective length of the optical cavity. The sixth term represents the piezomechanical interaction between microwave field and the mechanical phonon mode, where $g_{em}$ is the piezomechanical coupling strength. The sixth term represents the coupling between the QD and the mechanical phonon mode, $g$ being the corresponding coupling strength. The last three terms give the energy of the input fields. The intensities of the input microwave field, optical pump field and the optical probe field are described as $E_{m}=\sqrt{P_{m}\kappa_c/(\hbar\omega_{mi})}$,$E_{p}=\sqrt{P_{a}\kappa_a/(\hbar\omega_{ai})}$ and $E_{pr}=\sqrt{P_{r}\kappa_a/(\hbar\omega_{pr})}$ respectively, with $P_{m}$, $P_{a}$ and $P_{r}$ are the corresponding input powers.    \\
Using the standard linearization process under strong microwave driving field, the Hamiltonian in equation (1) can be rewritten as-
	
	\begin{eqnarray}\label{equation1}
H&=& \hbar\Delta_aa^{\dagger}a+\hbar\Delta_cc^{\dagger}c+\hbar\omega_{b}b^{\dagger}b+(\hbar/2)\omega_{q}\sigma_{z}-\hbar{g_{om}}a^{\dagger}a(b+b^{\dagger})-\hbar{G_{em}}(c^{\dagger}b+b^{\dagger}c)
\nonumber\\
&+&\hbar{g}(b^{\dagger}\sigma_{-}+b\sigma_{+})+\iota\hbar{E_p}(a^{\dagger}-a)+\iota\hbar{E_m}(c^{\dagger}-c)+\iota\hbar{E_{pr}}(a^{\dagger}e^{-\iota\delta{t}}-a{e^{\iota\delta{t}}})
\end{eqnarray}	
where, $\Delta_c=\delta_c-\frac{g_{em}^{2}E_m{^2}}{\omega_b(\kappa_c{^2}+\Delta_c{^2})}$, $\Delta_a=\delta_a+\frac{g_{om}g_{em}E_m{^2}}{\omega_b(\kappa_c{^2}+\Delta_c{^2})}$ and $G_{em}=g_{em}\sqrt{\frac{E_m^{2}}{(\kappa_c{^2}+\Delta_c{^2})}}$.\\

	The dynamics of the system can be described by quantum Langevin equations (QLE) including noise and damping terms. The nonlinear QLEs can be written as \\
	\begin{eqnarray}\label{equation2(a)}
	\dot{a}= -[\iota\Delta_a+\kappa_a/2]a+\iota{g_{om}}a(b^{\dagger}+b)+E_p+E_{pr}e^{-\iota\delta{t}}+A_{in}
\end{eqnarray}
	\begin{eqnarray}\label{equation2(b)}	
	\dot{b}= -[\iota\omega_b+\gamma_b/2]b+\iota{g_{om}}(a^{\dagger}a)+\iota{G_{em}}c-\iota{g}\sigma_-+B_{in}
	\end{eqnarray} 	
	\begin{eqnarray}\label{equation2(c)}
	\dot{c}= -[\iota\Delta_c+\kappa_c/2]c+\iota{G_{em}}b+E_m+C_{in}
	\end{eqnarray} 
	\begin{eqnarray}\label{equation2(d)}
	\dot{\sigma_-}= -[\iota\omega_q/2+\gamma_d/2]\sigma_-+2\iota{g}b\sigma_z+\Sigma_{in}
	\end{eqnarray} 
	
	The system is also interacting with the external degrees of freedom and therefore we introduce $\kappa_a$, $\gamma_b$, $\kappa_c$ and $\gamma_d$ as the decay constants of cavity field, moving mirror, microwave field and the QD respectively. $A_{in}$ and $C_{in}$ represents the optical and microwave input noises respectively; $C_{in}$ is the quantum Brownian noise associated with the movable mirror and $\Sigma_{in}$ is the noise related to QD.\\
	
	\section{Controllable optical bistability: Optical switching}
	
	We are interested in the steady state solutions of the equations (3)-(6). We take $a_s$, $b_s$, $c_s$ and $\sigma_{-s}$ as the average values of  the operators $a$, $b$, $c$ and $\sigma_{-}$ respectively under the conditions of strong microwave and optical fields i.e, $E_m>>E_{pr}$ and $E_p>>E_{pr}$. The steady state equations so obtained after putting time derivative to zero are-
	 	
	\begin{eqnarray}\label{equation3}
	a_s = \frac{E_p}{\iota\Delta^{'}_{a}+\kappa_a/2}
\nonumber\\
	b_s = \frac{\iota{g_{om}}|a_s|^2+\iota{G_{em}}c_s-\iota{g}\sigma_{-s}}{\iota\omega_b+\gamma_b/2}
	\nonumber\\
	c_s = \frac{\iota{G_{em}}b_s+E_m}{\iota\Delta_c+\kappa_c/2}
	\nonumber\\
	\sigma_{-s} = \frac{4\iota{g}<\sigma_z>b_s}{\gamma_d+\iota\omega_q}
	\end{eqnarray} 
	here $\Delta^{'}_{a}=\Delta_a-g_{om}(b_s+b^{*}_{s})$ is the effective detuning between optical cavity and the optical pump field due to mechanical motion of the movable mirror.\\
	
	We obtain the mean number of photons $|a_s|^{2}$ from equation (7) as
	
	\begin{eqnarray}\label{equation3}
	|a_s|^{2} = \frac{E_p^{2}}{\kappa_a^{2}/4+A_6^2+A_{7}^2A_5^{2}-2A_5A_6A_7}
	\end{eqnarray} 
	
	Where  $A_1=\gamma_b/2+\frac{G_{em}^2\kappa_c}{2(\Delta_c^{2}+\kappa_c^{2}/4)}-\frac{4 g^2\sigma_z\omega_q}{\omega_q^{2}+\gamma_d^{2}}$,     $A_2=\omega_b-\frac{G_{em}^2\Delta_c}{\Delta_c^{2}+\kappa_c^{2}/4}+\frac{4 g^2\sigma_z\omega_q}{\omega_q^{2}+\gamma_d^{2}}$,    $A_3=\frac{\kappa_cG_{em}E_m}{2(\Delta_c^{2}+\kappa_c^{2}/4)}$,  $A_4=\frac{\Delta_cG_{em}E_m}{\Delta_c^{2}+\kappa_c^{2}/4}$, $A_5=g_{om}|a_s|^{2}+A_3$, $A_6=\Delta_a-\frac{2g_{om}A_1A_4}{A_1^{2}+A_2^{2}}$, $A_7=\frac{2g_{om}A_2}{A_1^{2}+A_2^{2}}$,\\

	In order to quantify our study, we consider experimentally realizable values of the various parameters. In our study of the steady state and NMS, we will work in the resolved side band regime in which $\omega_{b}>\kappa_{a}, \kappa_{b}$ but while analyzing OMIT, we will work in both the resolved side band regime as well as the case $\omega_{b}<\kappa_{a}, \omega_{b}>\kappa_{b}$. The later case does not meet the condition of resolved side band regime but the system is nearly sideband resolved  \citep{wu}. 
	
	  Optical bistability is essential when it comes to designing of all-optical switching devices\citep{prakash}. Fig.2 illustrates the optical switching behavior for different combinations of the three interaction parameters,  $G_{em}$, $g_{om}$ and $g$. As observed the dashed and thick line plot displays the typical optical switching characteristics since at a certain value of the input pump power $E_{p}$, the intracavity photon value $|a_{s}|^{2}$ jumps from a low to high value. The thick line plot displays this optical switching characteristic at a lower value of  $E_{p}$. The switching ratio is defined as the ratio of the maximum to minimum value of $|a_{s}|^{2}$. The switching ratio for the thick plot is 1.84 while that for the dashed curve is 2.21. Consequently one has to optimize between high switching value and low input power. On the other hand, for some specific combination of $G_{em}$, $g_{om}$ and $g$, the switching property is absent as evident from the thin line plot of Fig.2. For designing of optoelectronic devices, it is an advantage if the bistable behaviour occurs at low values of the input power $E_{p}$. We thus show that the system can exhibit optical switching behavior at low values of the input power by tuning the various interactions.  This controllable bistable behavior shows that the system can be used as all-optical switch, logic gates and memory device for quantum information processing where low energy power input is essential.
	
	\begin{figure}[ht]
\hspace{-0.0cm}
\begin{tabular}{c}
\includegraphics [scale=0.8] {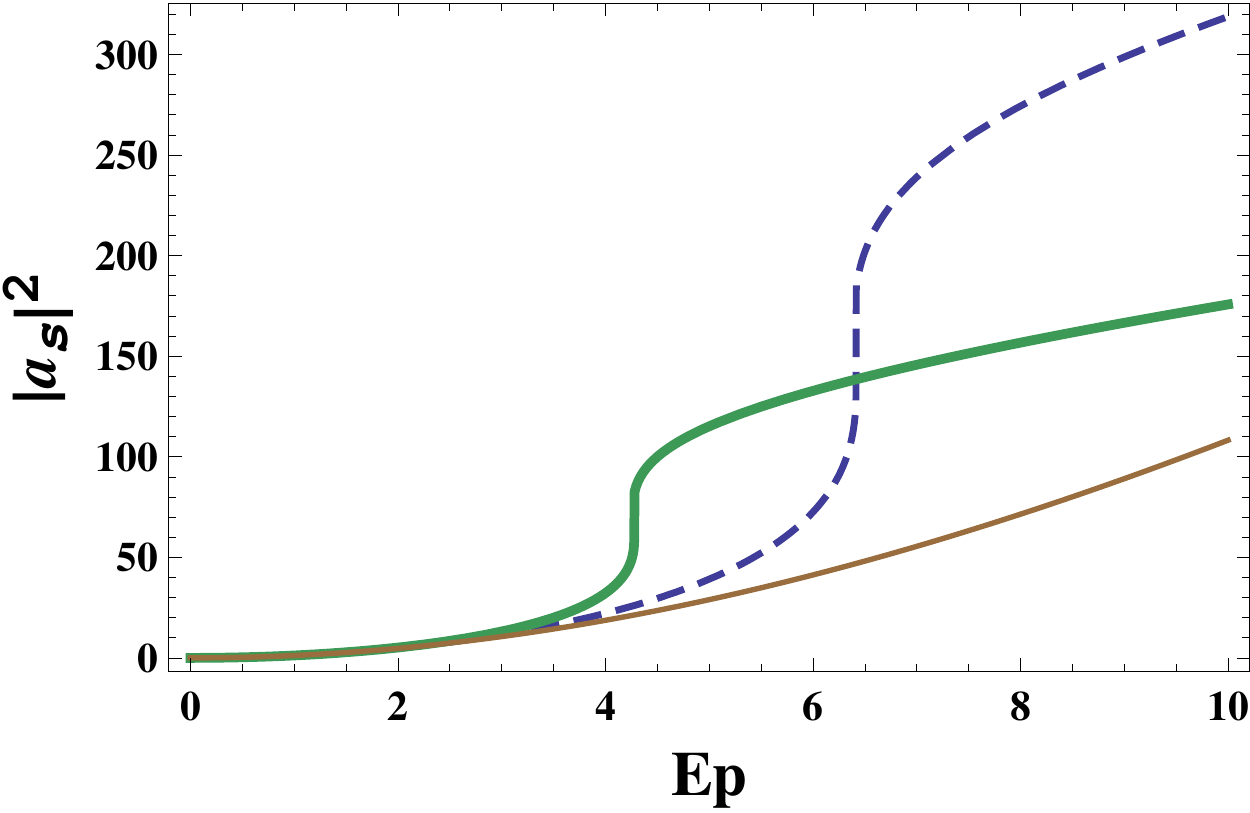} \\
\end{tabular} 
\caption{(Color online)The plot of steady state photon number as a function of pump power for different values of interaction parameters. (Dashed line): $G_{em}=0.1$,  $g_{om}=0.04$ and $g=0.001$, (Thick Solid Line): $G_{em}=0.1$, $g_{om}=0.06$ and $g=0.001$, (Thin Solid Line): $G_{em}=0.1$, $g_{om}=0.04$ and $g=0.001$ The other parameters used are  $\gamma_d= 0.000042$, $\gamma_b =0.000042$, $\kappa_c= 0.0000125$ , $\kappa_a= 0.9$ and $\sigma_{z}=-1$. All parameters are dimensionless with respect to $\omega_{b}$.}
\end{figure}

\section{Optomechanically induced transparency (OMIT)}

	Similar to the phenomenon of electromagnetically induced transparency (EIT) observed in atomic systems, optomechanical systems demonstrate opto-mechanically induced transparency (OMIT). The transparency window observed in the output field is due to the destructive interference induced between photons excited through different pathways. In this section, we discuss the generation of such a transparency window due to the opto-mechanical interaction. In particular, we will look into the possibility of controlling the OMIT by tuning the various interactions present in the system. To this end, we study the output characteristics of the probe field in the presence of optical field-mirror interaction ($g_{om}$), microwave field-mirror interaction ($G_{em}$) and qubit-mirror interaction ($g$). 
	
\begin{figure}[ht]
\hspace{-0.0cm}
\begin{tabular}{cc}
\includegraphics [scale=0.5] {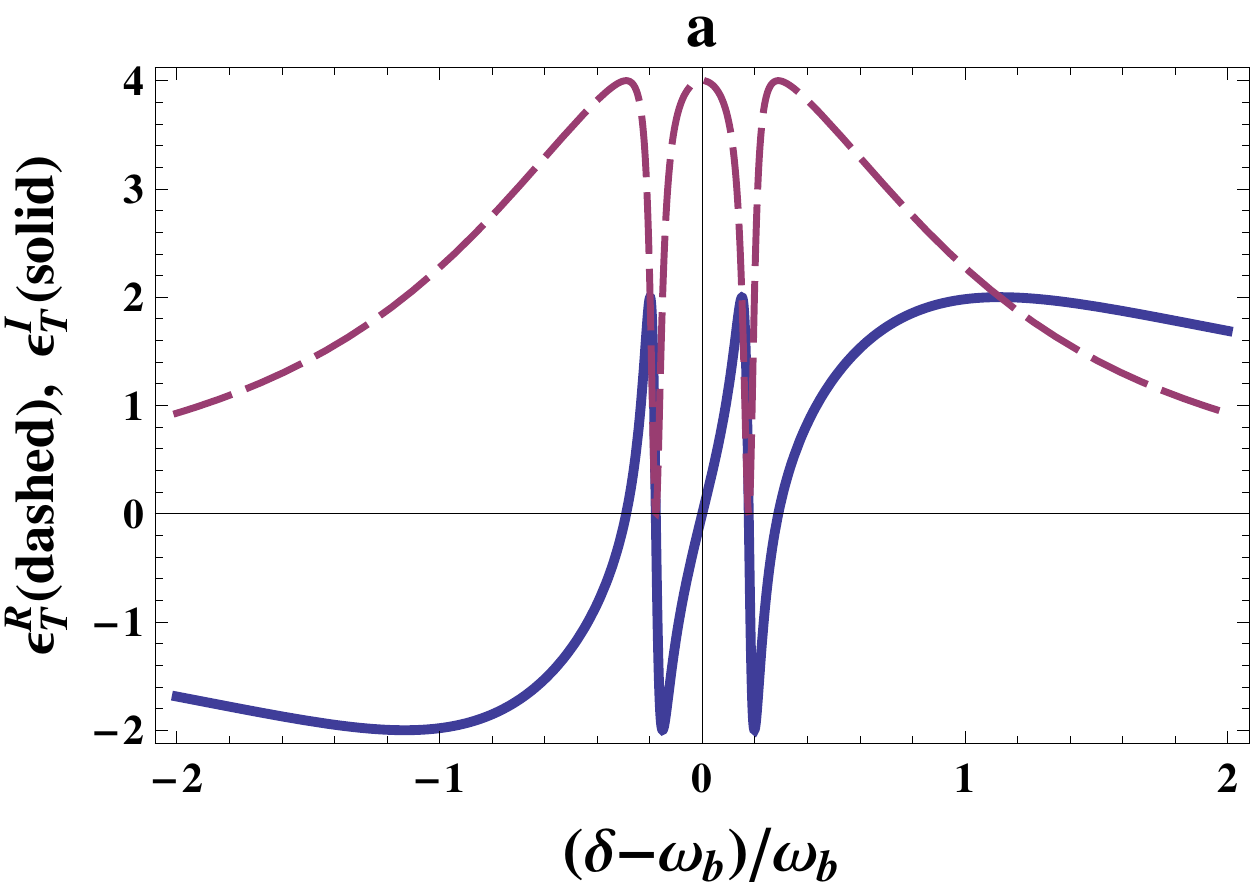} \includegraphics [scale=0.5] {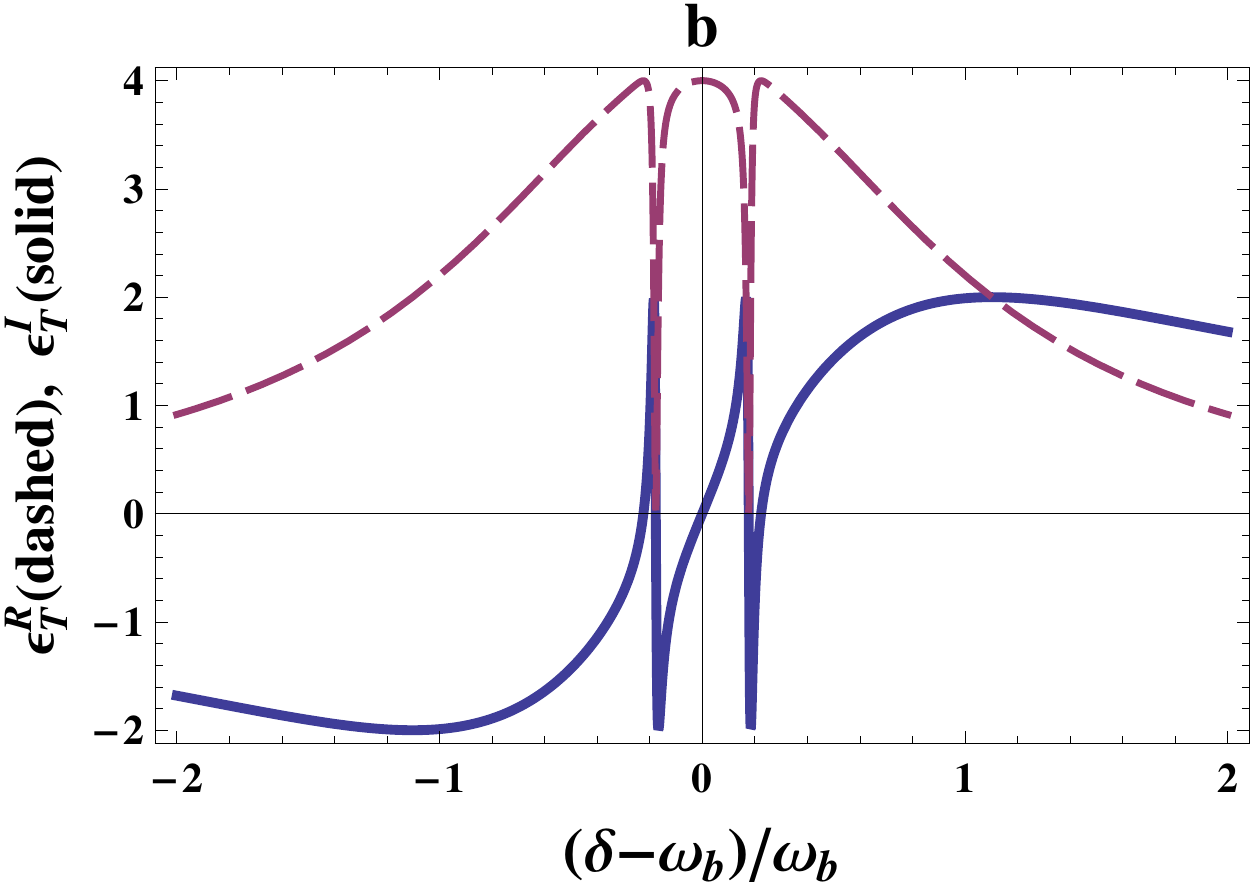}\\
\includegraphics [scale=0.5] {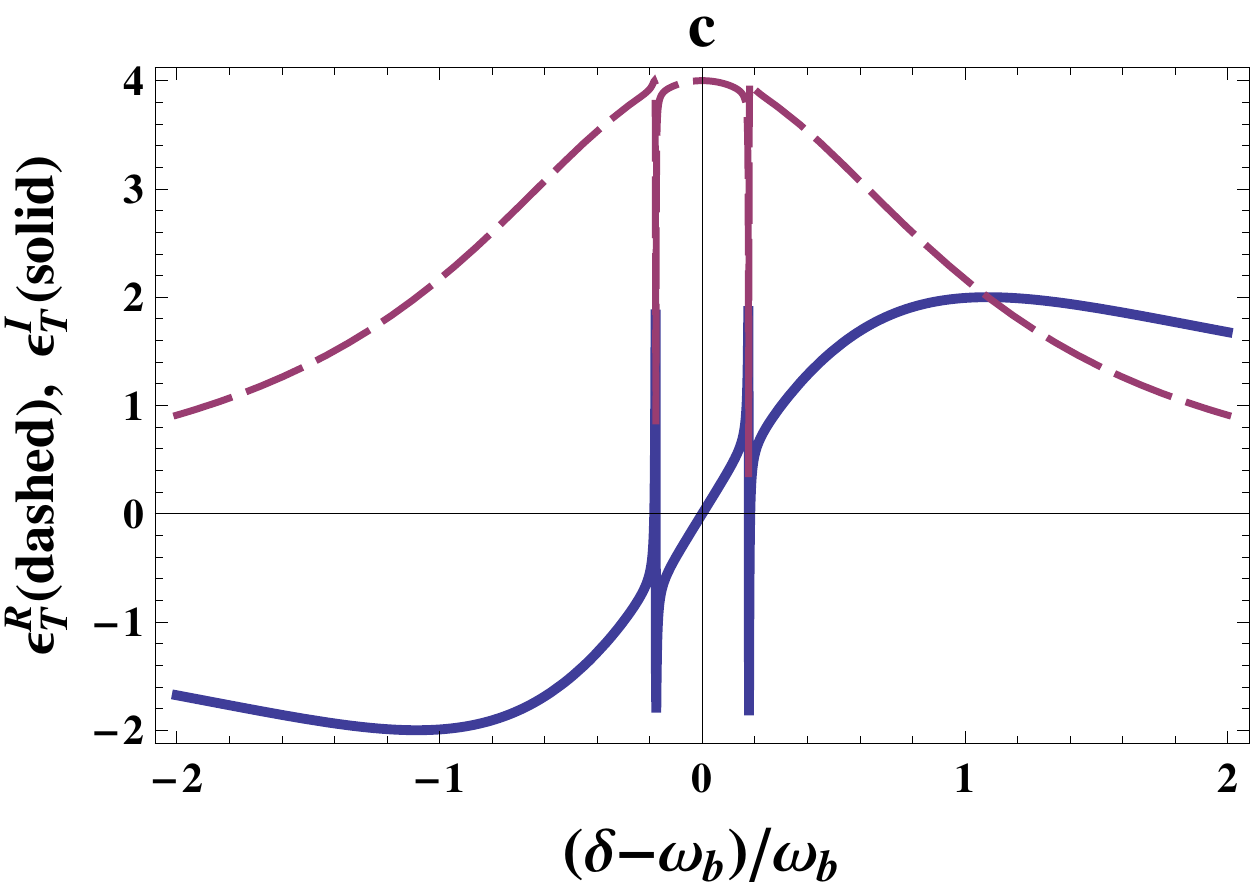} \includegraphics [scale=0.5] {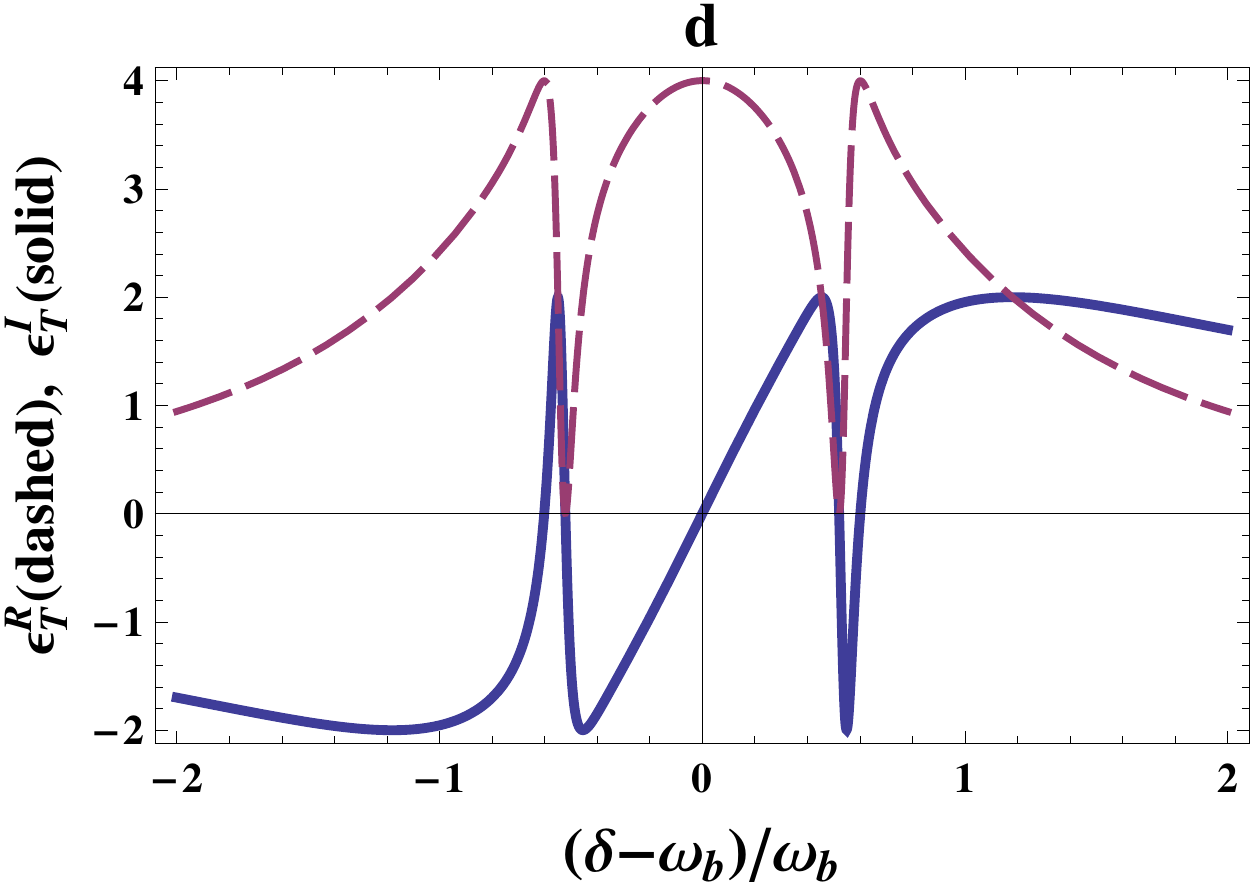}\\
\end{tabular} 
\caption{(Color online)The absorption Re[$\epsilon_T$] and dispersion Im[$\epsilon_T$ are plotted as a function of  $(\delta-\omega_b)/\omega_b$ for plot (a):$G_{om}=0.23$, $G_{em}=0.005$, $g=0.125$, (b): $G_{om}=0.1375$, $G_{em}=0.005$, $g=0.125$, (c)   $G_{om}=0.0458333$, $G_{em}=0.005$, $g=0.125$, (d):$G_{om}= G_{em} = g=0.3$ . The other parameters used are  $\gamma_d= 0.000042$, $\gamma_b =0.000042$, $\kappa_c= 0.0000125$ , $\kappa_a= 2.17$ and $\sigma_{z}=-1$. All parameters are dimensionless with respect to $\omega_{b}$.}
\end{figure}

\begin{figure}[ht]
\hspace{-0.0cm}
\begin{tabular}{cc}
\includegraphics [scale=0.5] {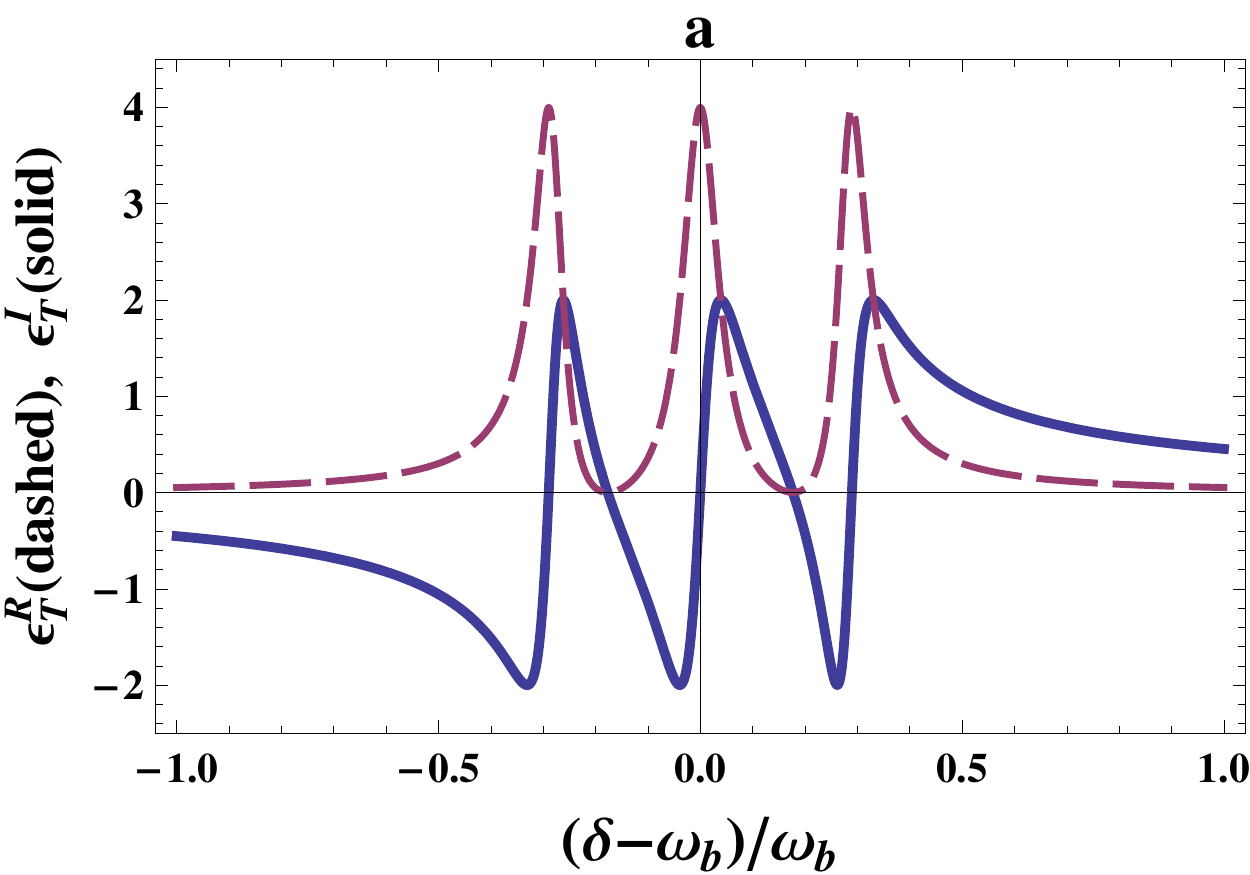} \includegraphics [scale=0.5] {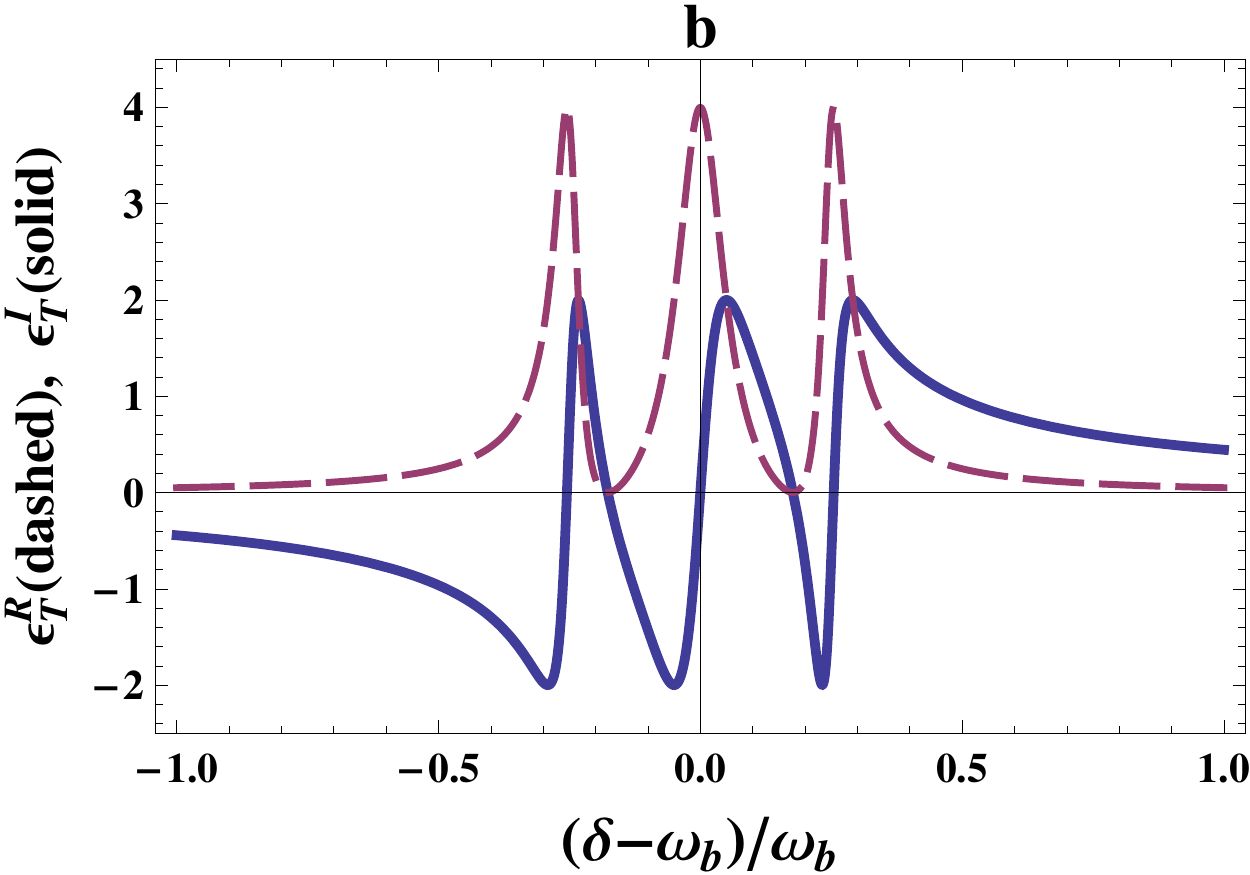}\\
\includegraphics [scale=0.5] {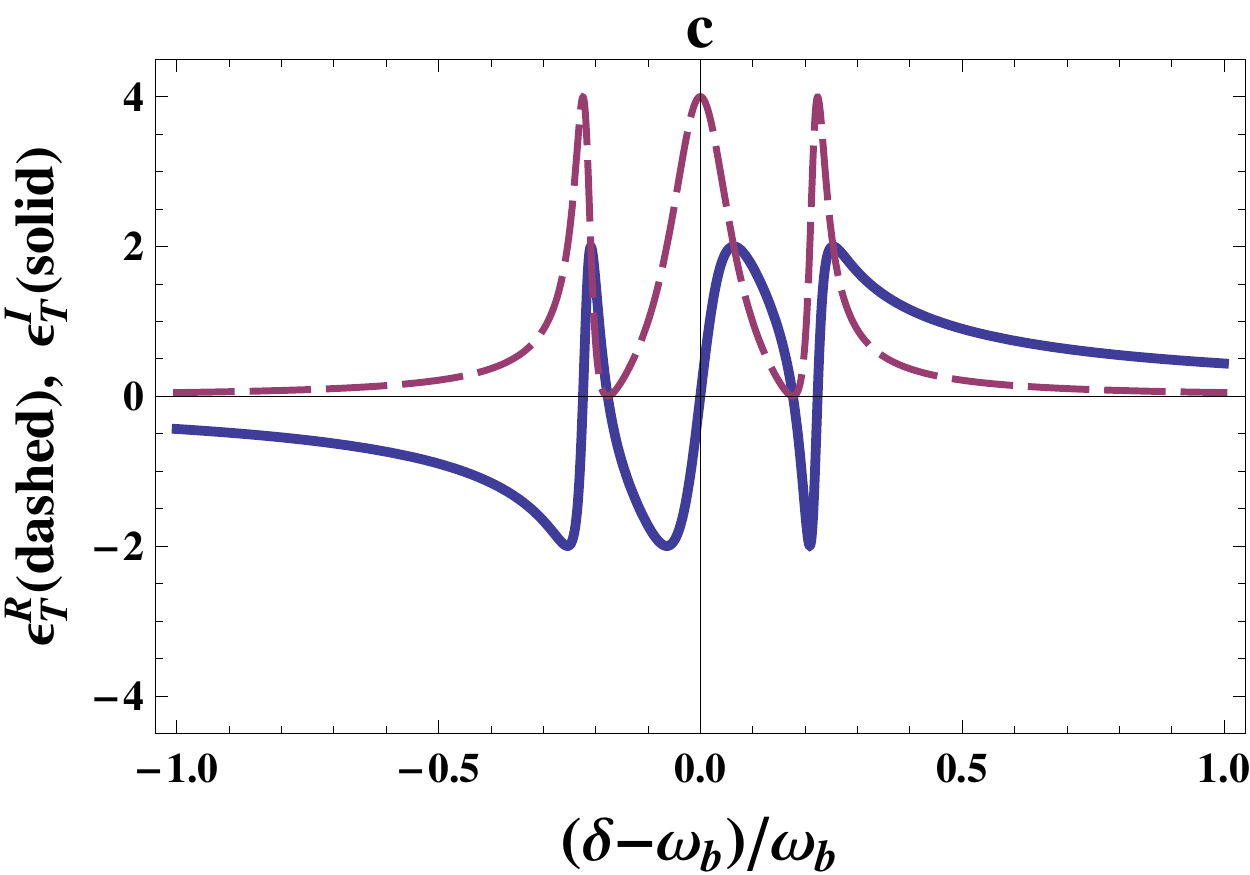} \includegraphics [scale=0.5] {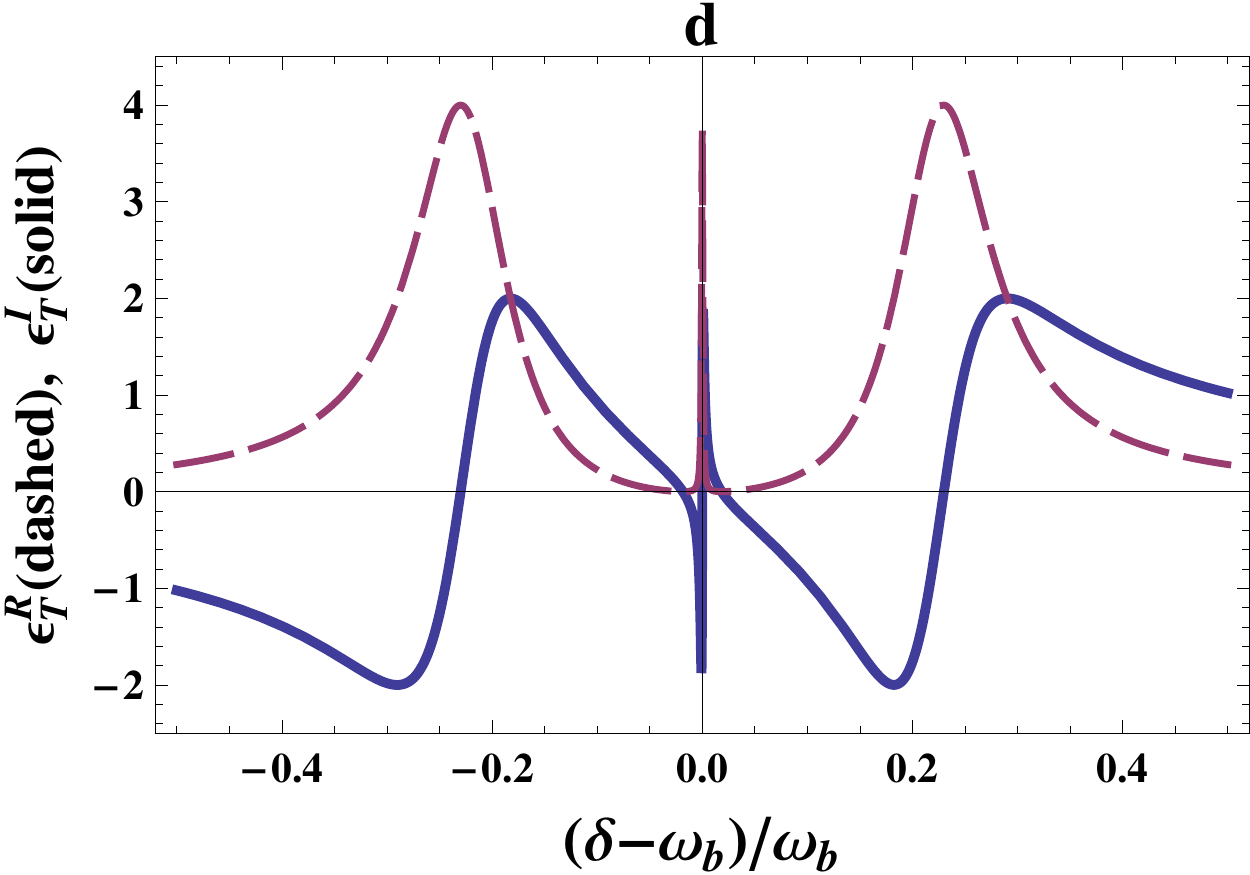}\\
\end{tabular} 
\caption{(Color online)The absorption Re[$\epsilon_T$] and dispersion Im[$\epsilon_T$ are plotted as a function of  $(\delta-\omega_b)/\omega_b$ for plot (a):$G_{om}=0.23$, $G_{em}=0.005$, $g=0.125$, (b) $G_{om}=0.183$, $G_{em}=0.005$, $g=0.125$, (c) $G_{om}=0.1375$, $G_{em}=0.005$, $g=0.125$, (d):$G_{om}=0.23$, $G_{em}=0.005$, $g=0.0125$ . The other parameters used are  $\gamma_d = 0.000042$, $\gamma_b =0.000042$, $\kappa_c= 0.0000125$ , $\kappa_a = 0.217$ and $\sigma_{z}=-1$. All parameters are dimensionless with respect to $\omega_{b}$.}
\end{figure}
	
	 Now to study the dynamics of quantum fluctuations, we linearize the quantum Langevin equations of the system around its steady state with the assumptions $a(t)\rightarrow{a_s}+\delta{a}$, $b(t)\rightarrow{b_s}+\delta{b}$, $c(t)\rightarrow{c_s}+\delta{c}$ and $\sigma_-(t)\rightarrow{\sigma_{-s}}+\delta{\sigma}_{-}$ and neglecting the small nonlinear fluctuation terms, the corresponding quantum Langevin equations become
	\begin{eqnarray}\label{equation4}
\delta\dot{a}=-(\iota\Delta^{'}_{a}+\kappa_a/2)\delta{a}+\iota{G_{om}}(\delta{b}+\delta{b}^{\dagger})+E_{pr}e^{-\iota\delta{t}}+A_{in},
\nonumber\\
\delta\dot{b}=-[\iota\omega_b+\gamma_b/2]\delta{b}+\iota(G_{om}^{*}\delta{a}+G_{om}\delta{a^{\dagger}})+\iota{G_{em}}\delta{c}-\iota{g}\delta\sigma_-+B_{in},
\nonumber\\
\delta\dot{c}=-[\iota\Delta_c+\kappa_c/2]\delta{c}+\iota{G_{em}}\delta{b}+C_{in},
\nonumber\\
\delta\dot{\sigma_-}=-[\iota\omega_q/2+\gamma_d/2]\delta\sigma_-+2\iota{g}<\sigma_z>\delta{b}+\Sigma_{in}.
\end{eqnarray}  
	
	Here $G_{om}=g_{om}a_s$ is the net coupling strength between the mechanical mode and the optical mode. We assume that the quality factor of the mechanical oscillator is high ($\omega_b>>\gamma_b$) and also the proposed system is operating in the resolved sideband regime i.e, $\omega_b>>\kappa_a,\kappa_c$. By assuming  \\
\begin{eqnarray}\label{equation1}
\delta{a}=\delta{a_+}e^{-\iota\delta{t}}+\delta{a_-}e^{\iota\delta{t}},
\nonumber\\
\delta{b}=\delta{b_+}e^{-\iota\delta{t}}+\delta{b_-}e^{\iota\delta{t}},
\nonumber\\
\delta{c}=\delta{c_+}e^{-\iota\delta{t}}+\delta{c_-}e^{\iota\delta{t}},
\nonumber\\
\delta{\sigma_-}=\delta{\sigma_{-+}}e^{-\iota\delta{t}}+\delta{\sigma_{--}}e^{\iota\delta{t}},
\nonumber\\
\delta{A_{in}}=\delta{A_{in+}}e^{-\iota\delta{t}}+\delta{A_{in-}}e^{\iota\delta{t}},
\nonumber\\
\delta{B_{in}}=\delta{B_{in+}}e^{-\iota\delta{t}}+\delta{B_{in-}}e^{\iota\delta{t}},
\nonumber\\
\delta{C_{in}}=\delta{C_{in+}}e^{-\iota\delta{t}}+\delta{C_{in-}}e^{\iota\delta{t}},
\nonumber\\
\delta{\Sigma_{in}}=\delta{\Sigma_{in+}}e^{-\iota\delta{t}}+\delta{\Sigma_{in-}}e^{\iota\delta{t}}.
\end{eqnarray}

Here the plus component ($\delta{s_{+}}$, s=a,b,c,$\sigma_-$...) corresponds to the original frequency $\omega_{pr}$ and the minus component($\delta{s_{-}}$, s=a,b,c,$\sigma_-$...) corresponds to the frequency $2\omega_{ai}-\omega_{pr}$. Substituting equation (10) in equation (9), neglecting the second and higher order terms and equating the coefficients with same frequency ($\omega_{pr}$) we get \\
\begin{eqnarray}\label{equation4}
\delta\dot{a}_{+}=(\iota\lambda_{a}-\kappa_a/2)\delta{a_{+}}+\iota{G_{om}}\delta{b_+}+E_{pr}+\delta{A_{in+}}
\nonumber\\
\delta\dot{b}_{+}=(\iota\lambda_{b}-\gamma_b/2)\delta{b_{+}}+\iota{G_{om}^{*}}\delta{a_{+}}+\iota{G_{em}}\delta{c_{+}}-\iota{g}\delta\sigma_{-+}+\delta{B_{in+}}
\nonumber\\
\delta\dot{c}_{+}=(\iota\lambda_{c}-\kappa_c/2)\delta{c_{+}}+\iota{G_{em}}\delta{b_+}+\delta{C_{in+}}
\nonumber\\
\delta\dot{\sigma}_{-+}=(\iota\lambda_{z}-\gamma_d/2)\delta{\sigma_{-+}}+2\iota{g}<\sigma_z>\delta{b_+}+\delta\Sigma_{in+}
\end{eqnarray} 
where $\lambda_a$=$\delta-\Delta^{'}_{a}$, $\lambda_b$=$\delta-\omega_b$, $\lambda_c$=$\delta-\Delta_{c}$ and $\lambda_z$=$\delta-\omega_{q}$. Now to neglect the expectation values of the noise operators, we assume that the system under consideration is working in mK regime, therefore we can have $<\delta{A_{in+}}>$=$<\delta{B_{in+}}>$=$<\delta{C_{in+}}>$=$<\delta{\Sigma_{in+}}>$=0. Also under mean field steady state conditions, the expectation values of $\delta\dot{a}_{+}$, $\delta\dot{b}_{+}$, $\delta\dot{c}_{+}$and $\delta\dot{\sigma}_{-+}$ are zero. Taking expectation values of the equation (11) and using the conditions mentioned above, we get
\begin{eqnarray}\label{equation4}
0=(\iota\lambda_{a}-\kappa_a/2)<\delta{a_{+}}>+\iota{G_{om}}<\delta{b_+}>+E_{pr}
\nonumber\\
0=(\iota\lambda_{b}-\gamma_b/2)<\delta{b_{+}}>+\iota{G_{om}^{*}}<\delta{a_{+}}>+\iota{G_{em}}<\delta{c_{+}}>-\iota{g}<\delta\sigma_{-+}>
\nonumber\\
0=(\iota\lambda_{c}-\kappa_c/2)<\delta{c_{+}}>+\iota{G_{em}}<\delta{b_+}>
\nonumber\\
0=(\iota\lambda_{z}-\gamma_d/2)<\delta{\sigma_{-+}}>+2\iota{g}<\sigma_z><\delta{b_+}>
\end{eqnarray} 
 equation (12) can be solved to get $<\delta{a_{+}}>$, which is
 
 \begin{eqnarray}\label{equation4}
<\delta{a_{+}>}=\frac{E_{pr}}{(\kappa_{a}/2-\iota\lambda_a)+\frac{|G_{om}^{2}|}{(\gamma_b/2-\iota\lambda_b)+\frac{G_{em}^{2}}{\kappa_c/2-\iota\lambda_c}-\frac{2g^{2}<\sigma_z>}{\gamma_d/2-\iota\lambda_z}}}
\end{eqnarray} 

Using input-output formalism of the cavity, the output field at the original frequency $\omega_{pr}$ can be written as
 \begin{eqnarray}\label{equation4}
\epsilon_{out}=2\kappa_a<\delta{a_{+}>}-E_{pr}
\end{eqnarray}
Also the transmission coefficient of the probe field can be expressed as
\begin{eqnarray}\label{equation4}
T_{pr}=\frac{\epsilon_{out}}{E_{pr}}=\frac{2\kappa_a<\delta{a_{+}>}}{E_{pr}}-1
\end{eqnarray}
If we define $\epsilon_T$ as $\epsilon_T=\frac{2\kappa_a<\delta{a_{+}>}}{E_{pr}}$, the quadrature $\epsilon_T$ at the original frequency $\omega_{pr}$, we obtain

\begin{eqnarray}\label{equation5}
\epsilon_T=\frac{2\kappa_a}{(\kappa_{a}/2-\iota\lambda_a)+\frac{|G_{om}^{2}|}{(\gamma_b/2-\iota\lambda_b)+\frac{G_{em}^{2}}{\kappa_c/2-\iota\lambda_c}-\frac{2g^{2}<\sigma_z>}{\gamma_d/2-\iota\lambda_z}}}
\end{eqnarray} 

The real and the imaginary parts of $\epsilon_T$ gives the absorption and the dispersion of the system.
\\
We assume that the optical cavity field , the qubit as well as the microwave and cavity field are driven at the mechanical red sideband i.e $\Delta_a^{'}$ = $\Delta_c$ = $\omega_q$ = $\omega_b$. Consequently, $\lambda_a$ = $\lambda_b$ = $\lambda_c$ = $\lambda_z$ = $\lambda$. We can rewrite equation (16) in a more intuitive form as,

\begin{equation}
    \epsilon_T = \frac{2\kappa_a}{(\frac{\kappa_a}{2} -\iota\lambda) + \frac{A_1}{\lambda_1 - \iota\lambda}+\frac{A_2}{\lambda_2 - \iota\lambda}+\frac{A_3}{\lambda_3 - \iota\lambda}},
\end{equation}
\vspace{2mm}

where $\lambda_1$, $\lambda_2$ and $\lambda_3$ are the roots of the cubic equation,

\begin{multline}
 x^3-x^2(\frac{\gamma_d+\gamma_b+\gamma_c}{2})-x(G_{em}^2 - 2g^2<\sigma_z>+(\frac{\gamma_d\gamma_b+\gamma_d\kappa_c+\gamma_b\kappa_c}{4})) + \\ (\frac{\gamma_b\kappa_c\gamma_d}{8}+ \frac{G_{em}^2\gamma_d}{2} - g^2\kappa_c<\sigma_z>)=0,   
\end{multline}

\vspace{2mm}
and $A_1$, $A_2$ and $A_3$ are defined as,
\begin{equation}
    A_1 = G_{om}^2\frac{[{\lambda_1(\frac{\kappa_c + \gamma_d}{2})}-\frac{\kappa_c\gamma_d}{4} - \lambda_1^2]}{(\lambda_1-\lambda_3)(\lambda_1-\lambda_2)}
\end{equation}

\begin{equation}
    A_2 = G_{om}^2\frac{[{-\lambda_2(\frac{\kappa_c + \gamma_d}{2})}+\frac{\kappa_c\gamma_d}{4} + \lambda_2^2]}{(\lambda_1-\lambda_2)(\lambda_2-\lambda_3)}
\end{equation}

\begin{equation}
    A_3 = G_{om}^2\frac{[{\lambda_3(\frac{\kappa_c + \gamma_d}{2})}-\frac{\kappa_c\gamma_d}{4} + \lambda_3^2]}{(\lambda_2-\lambda_3)(\lambda_1-\lambda_3)}
\end{equation}

In Figs (3) and (4), we plot absorption Re[$\epsilon_T$] and dispersion Im[$\epsilon_T$] for different values of QD-mechanical mode coupling strength (g), microwave-mechanical mode coupling strength ($G_{em}$) and net optomechanical coupling strength ($G_{om}$). The beat of the probe and the pump field leads to a time-varying radiation-pressure force with beat frequency $\delta$. At certain values of $\delta$, the mechanical resonator is driven resonantly. Consequently, sidebands of the optical field is generated due to mechanical oscillations. The position of the sidebands depends on the various interaction present in the system. In the presence of a strong pump field and resolved sideband (or nearly sideband resolved) limit, the frequency of the probe field coincides with the dominant sideband. This leads to destructive interference between the sideband and the probe field. Consequently, the destructive interference results in a transparency window due to the cancellation of the intracavity field.

In our study, we begin our discussion in the nearly sideband resolved regime ($w_b<\kappa_a, w_b>\kappa_b$) by first considering $G_{om} = 0.23$, $G_{em}=0.23$ and $g=0.125$ and plot the real (dashed line) and imaginary (solid line) parts of $\epsilon_T$ as a function of $\frac{(\delta-\omega_b)}{\omega_b}$ in Fig.3 (a). This plot shows two transparency windows at points which are determined by the roots $\lambda_{\iota}(\iota = 1,2,3)$ of eqn (18). Since $\gamma_d,\gamma_b,\kappa_c << 1,$ approximately the position of the two minima points are obtained as $\epsilon_{T}^R$ (minima) $\approx$ $\pm$ $\sqrt{G_{em}^2 - 2g^2<\sigma_z>} \approx \pm 0.177$ which is close to the numerically obtained result of $\pm  0.19$. At each of the transparency windows, anomalous dispersion (negative group velocity) is also observed. Negative group velocity corresponds to fast light propagation. Keeping  all parameters fixed, we now reduce $G_{om} = 0.138$ and the results are shown in Fig 3 (b). The two minima points are still located at the same points indicating that the minima points are independent of $G_{om}$. Clearly, the two transparency windows become narrow along with a comparatively steeper anomalous dispersion. A further narrowing of the transparency windows and steeping of the dispersion around the minima points is observed by reducing $G_{om} = 0.04$ (Fig 3 (c)). For the case $G_{om} = G_{em}= g = 0.3$ as shown in Fig 3 (d), the OMIT windows become wider and the separation between the two transparency windows is greatly enhanced. The position of the minima points from $\pm$ $\sqrt{G_{em}^2 - 2g^2<\sigma_z>}$ is $\pm 0.51$ and numerically obtained values are $\pm 0.56$.
A large anomalous dispersion enhances the radiation pressure force and hence the optomechanical coupling. This effect can assist in additional ground-state cooling of the mechanical resonator. Thus our system can be used to improve the sensitivity and bandwidth of the mechanical displacement measuement which has numerous applications in metrology and quantum information. The phenomenon of double-OMIT can be used as a logic gate in an quantum communication.

   In the resolved sideband regime $(\omega_b>>\kappa_a,\kappa_b)$, a three peak OMIA effect is observed in Fig.4. In between the three absorption peaks, two transparency windows exits within which anomalous dispersion is observed. The position of the two minima points are still at $\pm$ $\sqrt{G_{em}^2 - 2g^2<\sigma_z>}$. From eqn.(18)-(21), in the $\gamma_d,\gamma_b, \kappa_c << 1$, the position of the absorption peaks are at $\lambda_1$ = 0, $\lambda_{2,3}$ = $\pm$ $\frac{\sqrt{G_{om}^4 + \kappa_a^2(G_{em}^2+2g^2)}}{\kappa_a}$. These values agree well with the numerical values obtained in Fig.4. Decreasing $G_{om}$ while keeping $G_{em}$ and g fixed, the absorption peak at $\delta = \omega_b$ (i.e $\lambda_1 = 0$) widens and the separation between the three absorption peaks also decreases as evident from Figs.(b) and (c). At the same time, the dispersion curve also becomes distorted. The effect of decreasing qubit-mechanical mode g is shown in Fig.4(d). The absorption peaks at $\delta = \omega_b$ becomes extremely narrow while the other two side peaks broadens. The separation between the absorption peaks also enhances together with a highly distorted dispersion curve.

\section{Normal Mode Splitting}

In an optomechanical system, intermixing of fluctuations of different modes about their mean value leads to an important phenomena known as Normal Mode Splitting (NMS). This phenomenon is ubiquitous in both classical as well as in quantum physics. In this phenomena, due to strong coupling between different modes, energy exchange takes place on a time scale much faster than the decoherence time of every mode. The optomechanical NMS analyzed in this work, involves driving four parametrically coupled non-degenerate modes out of equilibrium. Here we study NMS in the resolved sideband regime and calculate the position quadrature of small fluctuations of the mechanical oscillator. To study NMS, we first transform the equations (3), (4), (5) and (6) into frequency domain and are then solved for the corresponding displacement spectrum, which in frequency domain is defined as-

\begin{eqnarray}\label{equation4}
S_{x}(\omega)=\frac{1}{4\pi}\int{d\Omega{e^{-\iota(\omega+\Omega)t}}<\delta{x(\omega)}\delta{x(\Omega)}+\delta{x(\Omega)}\delta{x(\omega)}}>
\end{eqnarray}  
In Fourier space, the displacement spectrum is obtained as-

\begin{eqnarray}\label{equation4}
S_{x}(\omega)=\frac{1}{d(\omega)d(-\omega)}[A_{25}(\omega)A_{25}(-\omega)]+A_{26}(\omega)A_{26}(-\omega)+A_{27}(\omega)A_{27}(-\omega)+\\ \nonumber
A_{28}(\omega)A_{28}(-\omega)+A_{29}(\omega)A_{29}(-\omega)+A_{30}(\omega)A_{30}(-\omega)+A_{32}(\omega)A_{32}(-\omega)+1]
\end{eqnarray}  

The various constants appearing in above equations have been defined in the appendix A

\begin{figure}[ht]
\hspace{-0.0cm}
\begin{tabular}{cc}
\includegraphics [scale=0.6] {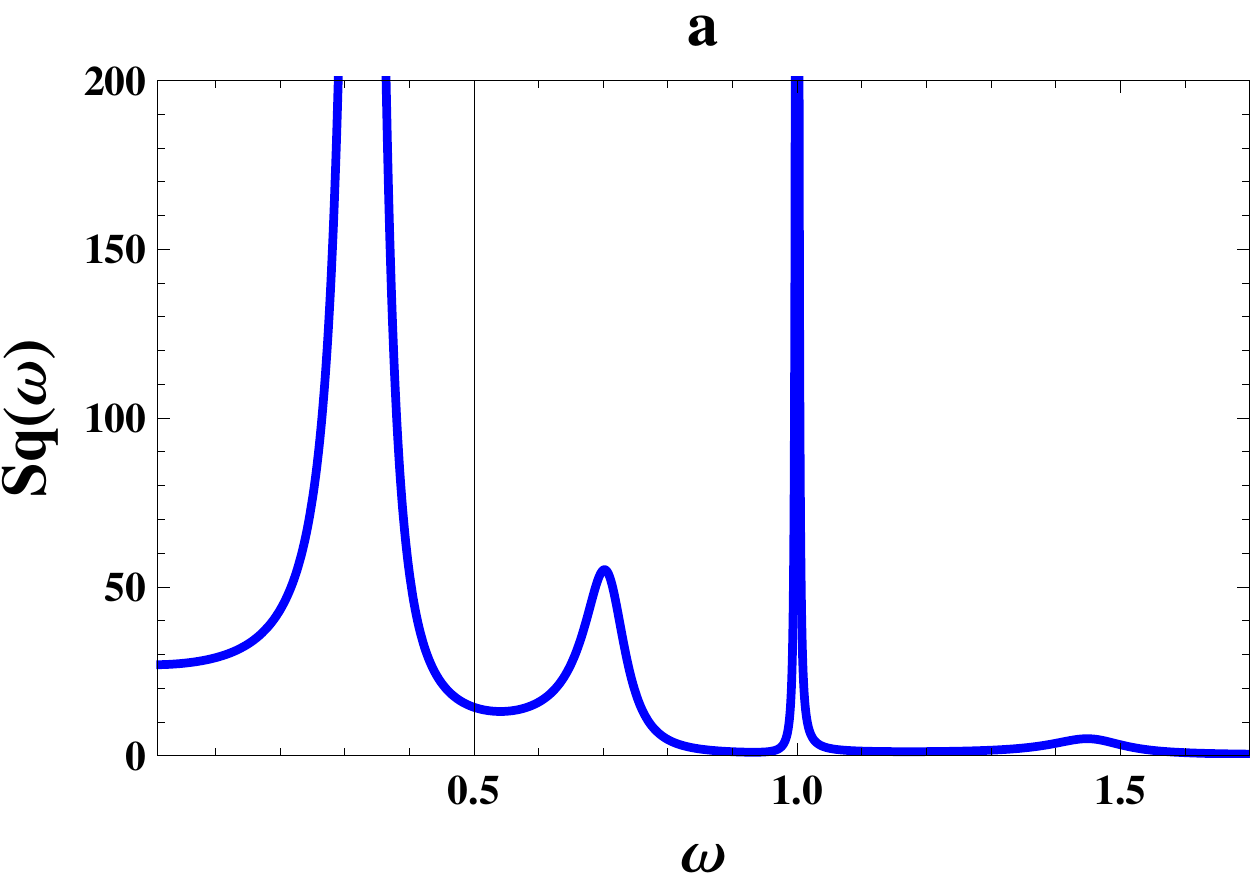} \includegraphics [scale=0.6] {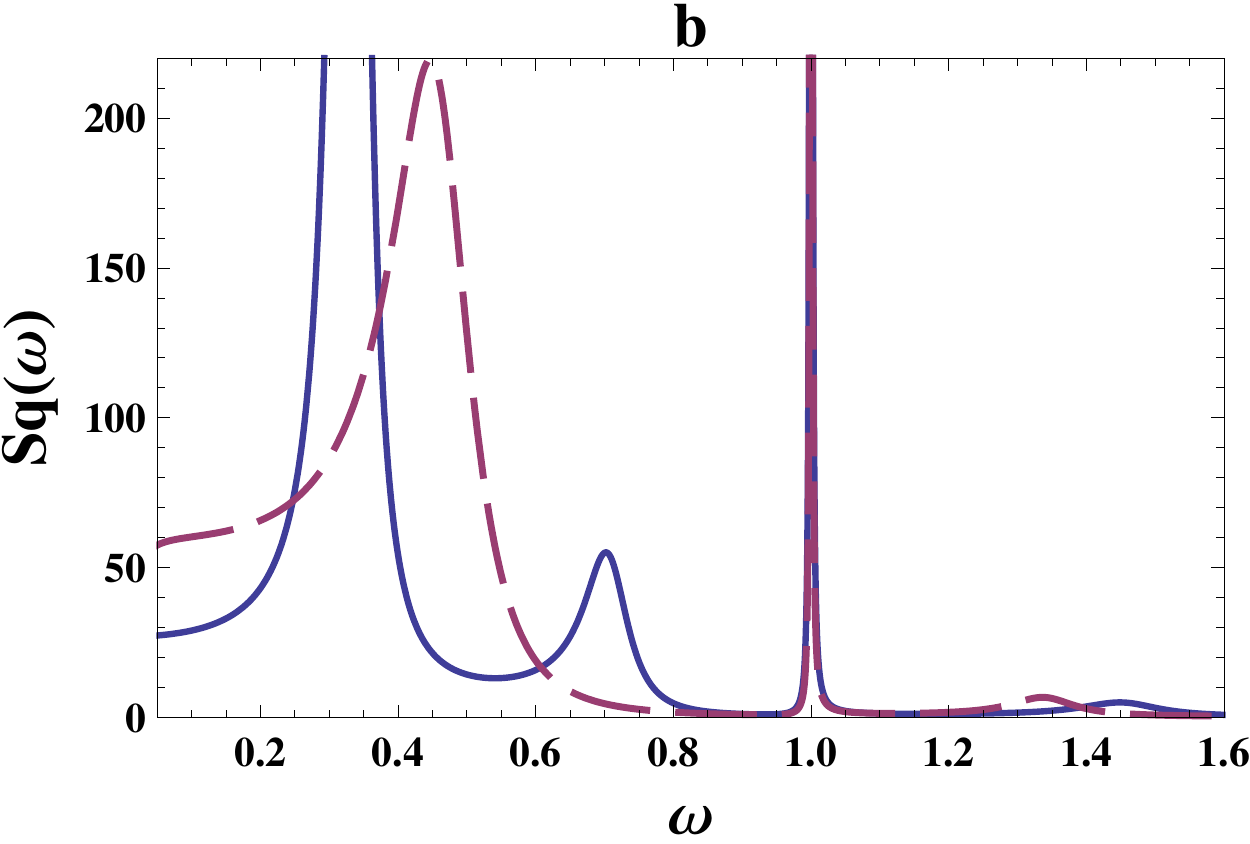}\\
\includegraphics [scale=0.6] {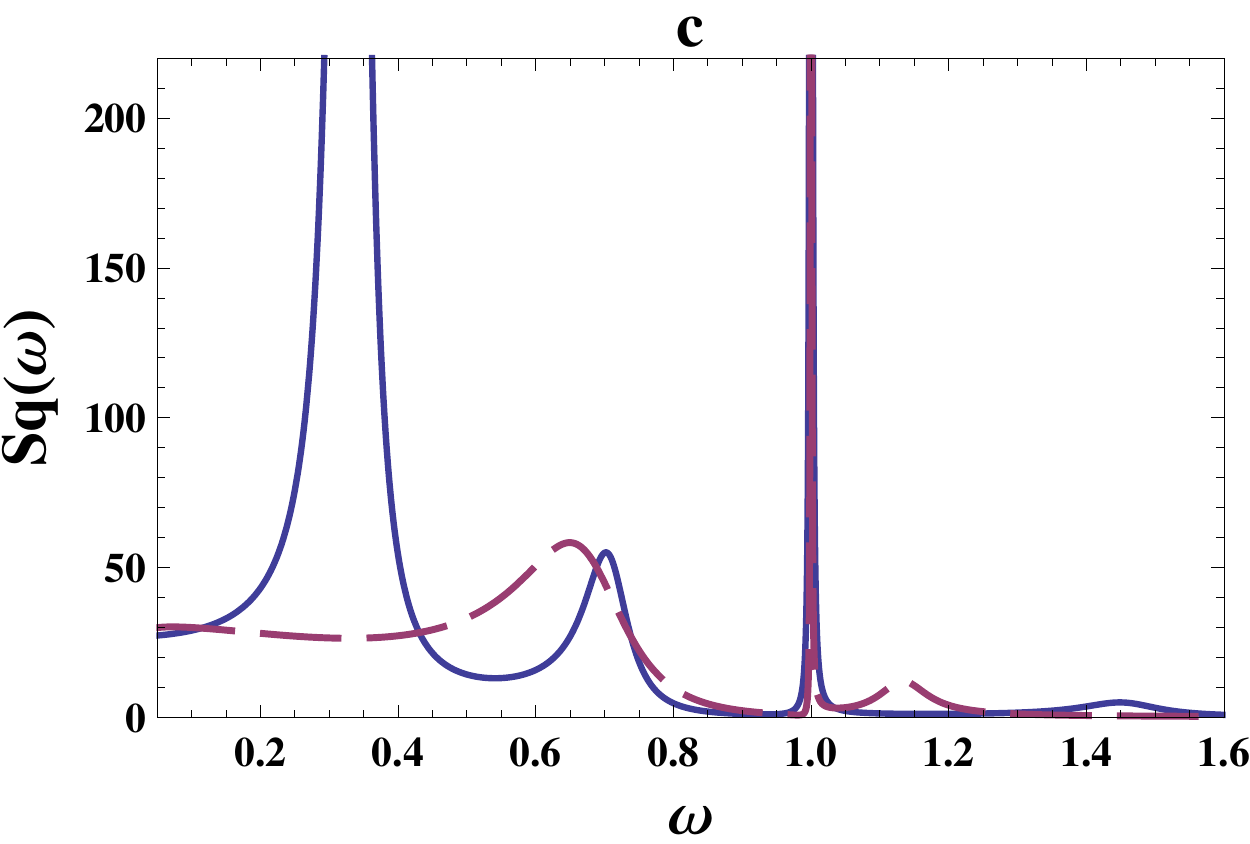} \includegraphics [scale=0.6] {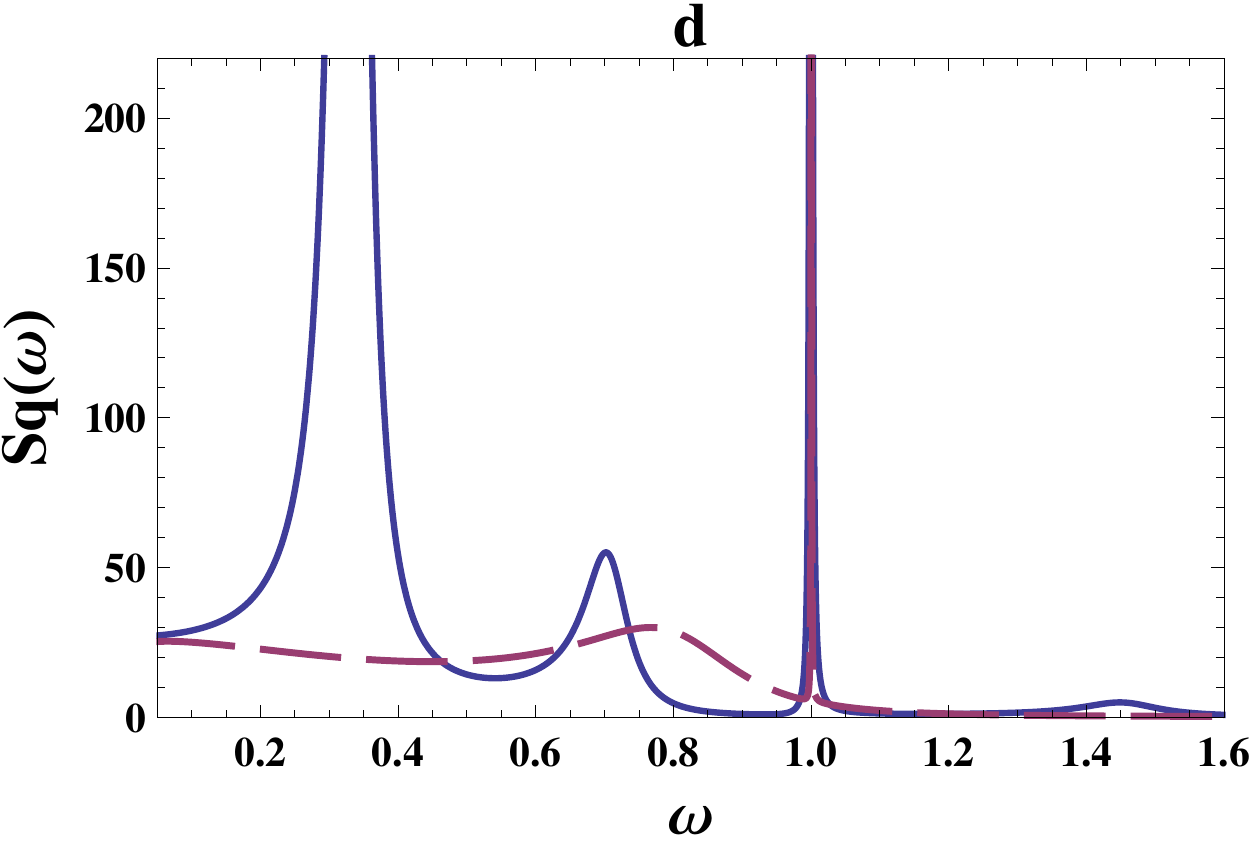}\\
\end{tabular} 
\caption{(Color online) Displacement spectrum of the mechanical resonator (Eqn.(18)) plotted as a function of  $\omega$ for plot (a) $G_{om}=G_{em}=g=0.4$, (b) Solid line: $G_{om}=G_{em}=g=0.4$, Dashed line:$G_{om}=G_{em}=0.4, g=0.01$  (c) Solid line:$G_{om}=G_{em}=g=0.4$, Dashed line: $G_{om}= 0.4, G_{em}= 0.2, g=0.01$ (d)  Solid line:$G_{om}=G_{em}=g=0.4$, Dashed line: $G_{om}= 0.4, G_{em}= 0.01, g=0.01$. The other parameters used are  $\gamma_d = 0.000042$, $\gamma_b =0.000042$, $\kappa_c= 0.0000125$ , $\kappa_a = 0.8$ and $\sigma_{z}=-1$. All parameters are dimensionless with respect to $\omega_{b}$.}
\end{figure}

In Fig.5, we plot the displacement spectrum $S_{x}(\omega)$ for different interactions $G_{om}$, $G_{em}$ and $g$. In Fig.5(a), the displacement spectrum for  $G_{om}=G_{em}=g=0.4$ is shown. We observe the NMS with four distinct peaks corresponding to all four modes of the system since all the three interactions coupling the four modes are equal. Keeping $G_{om}=G_{em}=0.4$ and reducing the QD-mechanical mode coupling $g=0.01$, the corresponding NMS (dashed curve) is displayed in Fig.5(b). In the limit of low value of $g$, the system reduces to the case of three mode coupling and hence the NMS displays three peaks. The peak near $\omega=1.45$ shifts to around $\omega=1.35$ and the peak at $\omega=0.7$ disappears along with a decrease in the height and shifting of the peak at  $\omega=0.3$. Reducing the piezomechanical coupling strength $G_{em}=0.2$ with $G_{om}=0.4$ and $g=0.01$, the corresponding NMS displayed (dashed curve) in Fig.4(c) shows a three peak structure with a drastically shifted and reduced peak intensity at $\omega=0.3$. Fig.4(d) illustrates (dashed curve) the influence of decreasing the interactions $G_{em}= g=0.01$ further and keeping opto-mechanical coupling strength high as $G_{om}=0.4$. The NMS reveals clearly a two peak structure corresponding to the mixing of the fluctuations of the optical and mechanical modes. Hence a selective energy exchange between the various modes of the system can be achieved by appropriately tuning the interactions.

We now discuss the experimental feasibility of the proposed system. In some recent experiments, optomechanical oscillators have been fabricated using piezoelectric materials such as AIN \citep{bochmann}. The AIN -nanobeam resonator can be driven simultaneously by both the microwave and optical fields under the effects of piezoelectric and radiation pressure interaction. It has been shown that the coupling strength between microwave and piezoelectric mechanical modes in a superconducting coplaner microwave cavity system can reach $12.3 \times 10^{6}$ Hz. \citep{clzou}. Based on realistic systems, the single-photon optomechanical coupling strength can exceed $g_{om}/ 2  \pi = 1.1$ MHz \citep{balram}. We have taken the frequency of the AIN-nanobeam resonator and the decay rate of the optical cavity as $\omega_{b}/2 \pi=2.4$ GHz and $\kappa_{a}/2 \pi = 5.2$ GHz (for the sideband resolved regime) and $\kappa_{a}/2 \pi = 0.5$ GHz (for resolved sideband regime). The quality factor of the microwave cavity is taken to be $2 \times 10^{5}$ \citep{megrant}. The qubit considered here could be an intrinsic defect inside the mechanical resonator or a superconducting circuit. Experimentally, a Jaynes-Cummings type of interaction between a superconducting qubit and a mechanical resonator was achieved \citep{cleland}. We have taken the qubit-mechanical oscillator interaction strength to be between $1-100$ MHz.

\section{conclusions}
In summary, we have investigated the optical response properties of a hybrid electro-optomechanical system in the presence of a qubit coupled to the mechanical oscillator via the Jaynes-Cummings interaction. The mean-field optical bistability analysis shows that the proposed system displays the typical, optical switching characteristics which can be tuned to function at low input power. The fluctuation dynamics reveals a series of interesting optical effects in the probe spectrum. In the sideband resolved regime ($\omega_{b}<\kappa_{a}$, $\omega_{b}>\kappa_{a}$), the absorption profile exibits double-OMIT while the dispersion profile shows negative group velocity (anomalous dispersion). On the other hand in the resolved sideband regime ($\omega_{b}>>\kappa_{a},\kappa_{b}$), a three peak OMIA effect is observed along with anomalous dispersion. Thus the system can be made to switch between OMIT and OMIA by tuning either the frequency of the mechanical oscillator or the optical cavity decay rate. Further the appearance of NMS shows tunable coherent energy exchange between the various sub-systems. These interesting optical properties are sensitive to the variations in $G_{om}$, $G_{em}$ and $g$. These experimentally feasible multi-transparency and multi-absorption phenomena provides the possibility for the realization of optical comb based on hybrid electro-optomechanical system interacting with a qubit. Our theoretical proposal provides a platform for novel quantum photonic devices.

\begin{acknowledgements}
		S. Yadav is grateful to BITS Pilani, Hyderabad campus for the PhD institute fellowship. Tarun Kumar thanks Ramjas College, University of Delhi for the facilities to carry out this research.
	\end{acknowledgements}

\section{Conflicts of interest/Competing interests}
The authors have no conflicts of interests.

\section{Authors' contributions}
A. B. Bhattacherjee proposed the model. All authors contributed equally to the analytical calculations, numerical simulations and discussing the physics.

\section{Appendix (A)}

\begin{eqnarray}\label{equation4}
\hspace{-1.0cm}
A_{25}(\omega)=\frac{2G_{om}\Delta_a^{'}}{d_1(\omega)}\lbrace\frac{(\omega_b+\frac{g^2\omega_z\sigma_z}{(-\iota\omega+\gamma_d/2)^2+\omega_z^{2}/4}-\frac{g_{em^2\Delta_c}}{(-\iota\omega+\kappa_c/2)^2+\Delta_c^{2}})g_{em}}{(-\iota\omega+\gamma_b/2-\frac{2g^2\sigma_z(-\iota\omega+\gamma_d/2)}{(-\iota\omega+\gamma_d/2)^2+\omega_z^{2}/4})(\frac{-\iota\omega+\kappa_c/2)^2+\Delta_c^{2}}{-\iota\omega+\kappa_c/2})+g_{em}^2}+\frac{\Delta_c}{-\iota\omega+\kappa_c/2}\rbrace
\end{eqnarray}

\begin{eqnarray}\label{equation4}
\hspace{-1.0cm}
A_{26}(\omega)=\frac{2G_{om}\Delta_a^{'}}{d_1(\omega)}\lbrace\frac{(\omega_b+\frac{g^2\omega_z\sigma_z}{(-\iota\omega+\gamma_d/2)^2+\omega_z^{2}/4}-\frac{g_{em^2\Delta_c}}{(-\iota\omega+\kappa_c/2)^2+\Delta_c^{2}})\Delta_cg_{em}}{(-\iota\omega+\gamma_b/2-\frac{2g^2\sigma_z(-\iota\omega+\gamma_d/2)}{(-\iota\omega+\gamma_d/2)^2+\omega_z^{2}/4})(-\iota\omega+\kappa_c/2)^2+\Delta_c^{2})+g_{em}^2(-\iota\omega+\kappa_c/2)}-1\rbrace
\nonumber\\
\end{eqnarray}

\begin{eqnarray}\label{equation4}
\hspace{-1.0cm}
A_{27}(\omega)=\frac{2G_{om}\Delta_a^{'}}{d_1(\omega)}\lbrace(\frac{g(-\iota\omega+\gamma_d/2)}{(-\iota\omega+\gamma_d/2)^2+\omega_{z}^2/4})-\frac{(\omega_b+\frac{g^2\omega_z\sigma_z}{(-\iota\omega+\gamma_d/2)^2+\omega_{z}^2/4}-\frac{g_{em}^2\Delta_c}{(-\iota\omega+\kappa_c/2)^2+\Delta_{c}^2})\frac{g\omega_z}{2(-\iota\omega+\gamma_d/2)^2+\omega_{z}^2/4}}{-\iota\omega+\gamma_b/2-\frac{2g^2\sigma_z(-\iota\omega+\gamma_d/2)}{(-\iota\omega+\gamma_d/2)^2+\omega_{z}^2/4}+\frac{g_{em}^2(-\iota\omega+\kappa_c/2)}{(-\iota\omega+\kappa_c/2)^2+\Delta_{c}^2}}\rbrace
\nonumber\\
\end{eqnarray}

\begin{eqnarray}\label{equation4}
\hspace{-1.0cm}
A_{28}(\omega)=\frac{2G_{om}\Delta_a^{'}g(-\iota\omega+\gamma_d/2)}{d_1(\omega)((-\iota\omega+\gamma_d/2)^2+\omega_{z}^2/4)}\lbrace\frac{\omega_z}{2(-\iota\omega+\gamma_d/2)}-\frac{(\omega_b+\frac{g^2\omega_z\sigma_z}{(-\iota\omega+\gamma_d/2)^2+\omega_{z}^2/4}-\frac{g_{em}^2\Delta_c}{(-\iota\omega+\kappa_c/2)^2+\Delta_{c}^2})}{-\iota\omega+\gamma_b/2-\frac{2g^2\sigma_z(-\iota\omega+\gamma_d/2)}{(-\iota\omega+\gamma_d/2)^2+\omega_{z}^2/4}+\frac{g_{em}^2(-\iota\omega+\kappa_c/2)}{(-\iota\omega+\kappa_c/2)^2+\Delta_{c}^2}}\rbrace
\nonumber\\
\end{eqnarray}

\begin{eqnarray}\label{equation4}
\hspace{-1.0cm}
A_{29}(\omega)=\frac{2G_{om}\Delta_a^{'}}{d_1(\omega)}\lbrace\frac{(\omega_b+\frac{g^2\omega_z\sigma_z}{(-\iota\omega+\gamma_d/2)^2+\omega_{z}^2/4}-\frac{g_{em}^2\Delta_c}{(-\iota\omega+\kappa_c/2)^2+\Delta_{c}^2})}{-\iota\omega+\gamma_b/2-\frac{2g^2\sigma_z(-\iota\omega+\gamma_d/2)}{(-\iota\omega+\gamma_d/2)^2+\omega_{z}^2/4}+\frac{g_{em}^2(-\iota\omega+\kappa_c/2)}{(-\iota\omega+\kappa_c/2)^2+\Delta_{c}^2}}\rbrace
\nonumber\\
\end{eqnarray}

\begin{eqnarray}\label{equation4}
\hspace{-1.0cm}
A_{30}(\omega)=\frac{2G_{om}\Delta_a^{'}}{d_1(\omega)}
\end{eqnarray}

\begin{eqnarray}\label{equation4}
\hspace{-1.0cm}
A_{32}(\omega)=\frac{2G_{om}\Delta_a^{'}}{d_1(\omega)(-\iota\omega+\kappa_a/2)}
\end{eqnarray}

\begin{eqnarray}\label{equation4}
\hspace{-1.0cm}
d_{1}(\omega)=\frac{-\iota\omega+\kappa_c/2}{-\iota\omega+\gamma_b/2-\frac{2g^2\sigma_z(-\iota\omega+\gamma_d/2)}{(-\iota\omega+\gamma_d/2)^2+\omega_z^{2}/4}+\frac{g_{em}^2(-\iota\omega+\kappa_c/2)}{(-\iota\omega+\kappa_c/2)^2+\Delta_c^{2}}}\lbrace(-\iota\omega+\gamma_b/2-\frac{2g^2\sigma_z(-\iota\omega+\gamma_d/2)}{(-\iota\omega+\gamma_d/2)^2+\omega_z^{2}/4}+
\nonumber\\
\frac{g_{em}^2(-\iota\omega+\kappa_c/2)}{(-\iota\omega+\kappa_c/2)^2+\Delta_{c}^2})^2+(\omega_b+\frac{g^2\omega_z\sigma_z}{(-\iota\omega+\gamma_d/2)^2+\omega_{z}^2/4}-\frac{g_{em}^2\Delta_c^{2}}{(-\iota\omega+\kappa_c/2)^2+\Delta_{c}^2})^2\rbrace
\nonumber\\
\end{eqnarray}

\end{document}